\documentclass[12pt]{article}
\usepackage{epsf}

\def\footnoteitem(#1)#2{
\begin{list}{#1}{\labelwidth4.0mm \leftmargin7.0mm
\labelsep2.5mm \rightmargin7.0mm \parsep0.5ex plus0.2ex minus0.1ex
\itemsep0ex plus0.2ex }
\item #2
\end{list}
}

\def\secteq#1{ \setcounter{equation}{0}
\renewcommand{\theequation}{#1.\arabic{equation}} }

\begin{document}
\newcommand{\be}{\begin{equation}}
\newcommand{\ee}{\end{equation}}
\newcommand{\ba}{\begin{eqnarray}}
\newcommand{\ea}{\end{eqnarray}}

\newcommand{\cL}{{\cal L}}
\newcommand{\cM}{{\cal M}}
\newcommand{\Bt}{{\tilde B}}
\newcommand{\cO}{{\cal O}}
\newcommand{\cOt}{{\tilde\cO}}
\newcommand{\bt}{{\tilde\beta}}
\newcommand{\tr}{{\mbox{tr}\,}}
\newcommand{\str}{{\mbox{str}\,}}
\newcommand{\Exp}{{\mbox{exp}\,}}
\newcommand{\Mdot}{{\dot M}}
\newcommand{\Mbar}{{M_{VS}}}
\newcommand{\tb}{{\tilde\beta}}
\newcommand{\vp}{{\vec p}}
\newcommand{\hX}{{\hat X}}
\newcommand{\diag}{{\rm diag}}
\newcommand{\sbar}{{\overline{s}}}
\newcommand{\dbar}{{\overline{d}}}
\newcommand{\ubar}{{\overline{u}}}
\newcommand{\qbar}{{\overline{q}}}
\newcommand{\psibar}{{\overline{\psi}}}
\newcommand{\tu}{{\tilde u}}
\newcommand{\tub}{{\overline{\tu}}}
\newcommand{\td}{{\tilde d}}
\newcommand{\tdb}{{\overline{\td}}}
\newcommand{\ts}{{\tilde s}}
\newcommand{\tsb}{{\overline{\ts}}}
\newcommand{\ie}{{\it i.e.}}
\newcommand{\cf}{{\it cf.}}
\newcommand{\etc}{{\it etc.}}
\newcommand{\Nh}{{\hat N}}
\newcommand{\Real}{{\rm Re}}
\newcommand{\Imag}{{\rm Im}}
\newcommand{\eps}{{\varepsilon}}

\begin{titlepage}
\begin{flushright}
SISSA 87/02/EP\\
\end{flushright}
\vskip 0.5cm
\begin{center}
{
\large \bf On the effects of (partial) quenching on penguin contributions to 
\boldmath{$K\to\pi\pi$} }
\vskip1cm {
Maarten Golterman$^a$ and  Elisabetta Pallante$^b$ }\\
\vspace{.5cm} {\normalsize {\sl $^a$ Department of Physics and Astronomy,
 San Francisco State University,\\
 1600 Holloway Ave, San Francisco, CA 94132, USA\\
\vspace{.2cm} $^b$ SISSA, Via Beirut 2--4, I-34013 Trieste, Italy  }}\\
 \vspace{2.0cm} 

{\bf Abstract\\[10pt]} \parbox[t]{\textwidth}{{
Recently, we pointed out that chiral transformation properties of 
strong penguin
operators change in the transition from unquenched to (partially) quenched
QCD. As a consequence, new penguin-like
operators appear in the (partially) quenched theory, along with
 new low-energy constants, which should
 be interpreted as a quenching artifact.  
Here, we extend the analysis to the contribution of the new low-energy 
constants to the
$K^0\to\pi^+\pi^-$ amplitude, at leading order in chiral
perturbation theory, and for arbitrary (momentum non-conserving)
kinematics.  
Using these results, we provide a detailed discussion
of the intrinsic systematic error due to this (partial) quenching artifact.
We also give a simple recipe for the determination of the leading-order
low-energy constant parameterizing the new operators in the case of
strong $LR$ penguins.}} 

\end{center}
\vfill
\noindent $^a$ e-mail: {\em maarten@quark.sfsu.edu}  \\
\noindent $^b$ e-mail: {\em pallante@he.sissa.it}  \\
\end{titlepage}
\section{\large\bf Introduction}
%
%
%
A reliable calculation of long-distance contributions to non-leptonic
 kaon-decay
rates, and, in particular, to the CP-violating part parametrized by the
quantity $\eps'/\eps$ has been a longstanding challenge.
Ideally, one would expect such calculations to be in the domain of
lattice QCD, but in practice many theoretical and numerical 
difficulties
have made progress in this direction rather slow.  Recently however,
two lattice collaborations have reported on numerical results for both
the real and imaginary parts of $\Delta I=1/2$ and  $\Delta I=3/2$
$K\to\pi\pi$ matrix elements with a rather satisfactory control over 
statistical errors
\cite{cppacs,rbc}.  These lattice computations were done with the 
effective weak $\Delta S=1$ hamiltonian with three flavors, \ie\ with
the charm integrated out, and they were possible because of the use of 
lattice fermions with good chiral symmetry.  Both groups reported values
of $\eps'/\eps$ which are non-zero, and thus consistent with
the existence of direct CP violation, but of opposite sign 
(and comparable size) to the experimentally measured value.

While statistical errors for these lattice computations seem to be 
reasonably under control, this is not the case for a large class of systematic
errors, which will need to be studied further in the future.
One source of systematic error is the use of the quenched
approximation.  In a previous paper \cite{gp} we pointed out that,
in addition to the fact that quenched QCD is just not the same theory
as full QCD, an ambiguity arises in the {\em definition} 
of the quenched version
of penguin operators appearing in the $\Delta S=1$ effective weak hamiltonian.
The ambiguity originates in the difference of the chiral transformation
properties of penguin operators within the quenched and unquenched theories.
This implies that not only will quenched lattice results be hampered by the 
fact that we do not really know whether
quenched values of given matrix elements are close to
their real-world values, but, in the case of penguins, they also depend 
on which definition of the operators is chosen, since more than one definition
 is possible.

In fact, both lattice computations \cite{cppacs,rbc}
did not directly compute $K\to\pi\pi$ matrix
elements, but $K\to\pi$ (with $M_\pi=M_K$) and $K\to 0$
transition amplitudes, and used chiral perturbation theory (ChPT) 
to convert them into the desired
$K\to\pi\pi$ matrix elements \cite{cbetal}.  In ref.~\cite{gp} we
explained how the chiral properties of penguin operators change
in the transition to the (partially) quenched theory, and how,
in principle, more than one definition of a quenched penguin
operator is possible.  Using ChPT,
we traced how this affects $K\to\pi$ and $K\to 0$ matrix elements.
We restricted ourselves to $LR$ penguins (\ie\ $Q_5$ and $Q_6$),
because the effects in this case already appear at leading order
in ChPT, while they are a next-to-leading order effect for $LL$
penguins.  Because of the fact that the ambiguity is already present 
at leading order for matrix elements of $Q_{5,6}$, this may be
an important issue for $\eps'/\eps$ (while it is expected
to be less important for the real parts of $K\to\pi\pi$ amplitudes,
and thus the $\Delta I=1/2$ rule).

In this paper we extend our ChPT calculations to the effect of the ambiguity 
on $K\to\pi\pi$ matrix elements, again to leading chiral order.  
This is important
for two reasons. First, lattice computations may be done directly 
for $K\to\pi\pi$
matrix elements, and their chiral behavior needs to be
known in order to fit lattice results as a function of quark masses.
Since lattice computations are typically done with unphysical (\ie\
energy/momentum non-conserving) kinematics,
we present our results for the most general kinematics
possible, both in the quenched and partially quenched cases.  
Second,
once the complete (leading-order) ChPT expressions for $K\to 0$, $K\to\pi$ 
and $K\to\pi\pi$ matrix elements are available, it is possible to give
a more detailed discussion of the systematic error introduced by the
ambiguity in the definition of quenched penguin operators.

The paper is organized as follows.  In section 2, we review the main
observation of ref.~\cite{gp}.  We show how a $LR$ penguin, which transforms
in an irreducible representation (irrep) of $SU(3)_L\times SU(3)_R$, splits
into two operators in the (partially) quenched theory, with each of them 
transforming in a different irrep of the (partially) quenched chiral symmetry 
group.
One of these irreps corresponds ``naturally" to the single irrep of the 
unquenched theory, while the other irrep can be considered as ``new," and 
an artifact of quenching.
We give the ChPT realization of all relevant operators at leading and 
next-to-leading chiral order, introducing
new low-energy constants (LECs) which appear in correspondence to the new
irrep.  In section 3,
we present our results for $K^0\to\pi^+\pi^-$ penguin matrix elements to
leading order in ChPT, with general kinematics, and specialize
these results to physical (\ie\ energy-momentum conserving) kinematics.  
In section 4, 
we discuss different strategies available for using quenched
lattice results to estimate real-world $K\to\pi\pi$ matrix elements, and give
some numerical examples.
We provide a simple prescription for determining the leading-order LEC 
representing the new irrep in section 5, and section 6 contains our 
conclusions.
Some of this work has already been presented in ref.~\cite{gplat2002}.
\section{ \large\bf 
Review of \boldmath{$LR$} penguins in (partially) quenched QCD and ChPT}
\secteq{2}
A lagrangian definition for partially quenched QCD can be constructed
as explained in ref.~\cite{bgpq} (see also ref.~\cite{replica} for an 
alternative realization using the replica method).  
In addition to the valence quarks $q_{vi}$,
$i=u,d,s$, with masses $m_{vi}$, one introduces a separate set of sea 
quarks $q_{si}$, $i=1,\dots,N$, with masses $m_{si}$, and a set of
``ghost" quarks $q_{gi}$, $i=u,d,s$, with masses equal to those of
the valence quarks $m_{gi}=m_{vi}$ \cite{morel}.  Ghost quarks
are given bosonic statistics, such that the ghost-quark determinant
cancels the valence-quark determinant, thus leaving only the
sea-quark determinant present in the path integral.  Therefore, only
 sea quarks propagate in internal loops.  

Since partially quenched QCD thus contains more flavors than unquenched
QCD, its flavor symmetry group is larger than the QCD one.  The full chiral
symmetry group relevant for light meson physics is the graded extension of the 
ordinary chiral group
$SU(3+N|3)_L\times SU(3+N|3)_R$ \cite{bgpq}.  It is graded
because part of its elements transform fermions into bosons and
{\it vice versa}.  The quenched theory,
which has no sea quarks at all, corresponds to the special case $N=0$
\cite{bgq}.

We consider $LR$ penguin operators of the form
\be
Q_{penguin}=(\sbar d)_L(\ubar u+\dbar d+\sbar s)_R\ ,\label{penguin}
\ee
where 
\ba
(\qbar_1 q_2)_{L,R}&=&\qbar_1\gamma_\mu P_{L,R}q_2\ ,\label{def}\\
P_{L,R}&=&\frac{1}{2}(1\mp\gamma_5) \ ,\nonumber
\ea
and color contractions are not specified, so that $Q_{penguin}$ can 
represent both $Q_5$ and $Q_6$.  As already pointed out in ref.~\cite{gp},
the  $u$, $d$ and $s$ fields in eq.~(\ref{penguin}) 
represent valence quarks
in the partially quenched theory, and the penguin operator
can be decomposed as\footnote{In a theory with $K$ valence quarks, all
ratios $3/N$ get replaced by $K/N$.}
\ba
Q_{penguin}&=&
\frac{3}{N}\;\str(\Lambda\psi\psibar\gamma_\mu P_L)
\;\str(\psi\psibar\gamma_\mu P_R)+
\str(\Lambda\psi\psibar\gamma_\mu P_L)
\;\str(A\psi\psibar\gamma_\mu P_R)\ , \nonumber \\
&\equiv&\frac{3}{N}\;Q^{PQS}_{penguin}+Q^{PQA}_{penguin}\ ,
\label{pqdecomp}\\
A&=&\diag(1-\frac{3}{N},1-\frac{3}{N},1-\frac{3}{N},
-\frac{3}{N},\dots,-\frac{3}{N})\ , \label{a} \\
\Lambda_{ij}&=&\delta_{is}\delta_{jd}\ ,\nonumber
\ea
where the first $3$ (valence) entries of $A$ are equal to $1-{3}/{N}$,
and the next $N+3$ (sea and ghost) entries are equal to $-{3}/{N}$.
Here $\psi$ collects all quark fields in the theory,
$\psi=(q_{vi},q_{si},q_{gi})$.  The tensor $\Lambda$ projects onto the
valence $(\sbar d)_L$ term in the first factor of $Q_{penguin}$.
The motivation for splitting $Q_{penguin}$ this way is that 
$Q^{PQS}_{penguin}$ and $Q^{PQA}_{penguin}$ form different
representations of the partially quenched symmetry group:
$Q^{PQS}_{penguin}$ ($Q^{PQA}_{penguin}$) transforms in the
trivial (adjoint) irrep of $SU(3+N|3)_R$.
As a consequence, there are at least two different ways of embedding the
QCD penguin operator into the partially quenched theory.  One is
to choose the partially quenched penguin to be a singlet under
$SU(3+N|3)_R$, \ie\ $Q^{PQS}_{penguin}$, as in the unquenched theory, 
whereas the other choice is to use the original
operator, which is seen to be a linear combination of two
irreducible operators.  This latter choice was made in 
refs.~\cite{cppacs,rbc}. 
In ref.~\cite{gpchpt} non-singlet penguin operators such as
$Q^{PQA}_{penguin}$ were not considered, because singlet factors
such as $(\ubar u+\dbar d+\sbar s)_R$ in eq.~(\ref{penguin}) 
had been implicitly extended to singlets under the
full (partially-)quenched symmetry group.  Therefore, the analysis
of ref.~\cite{gpchpt} was not complete, and ref.~\cite{gp} and this
paper remedy this for LR penguins.

To leading order (LO), the operators $Q^{PQS}_{penguin}$, $Q^{PQA}_{penguin}$
are represented in ChPT by \cite{gp}
\ba
Q^{PQS}_{penguin}&\rightarrow& -\alpha^{(8,1)}_1\;\str(\Lambda L_\mu L_\mu)
+\alpha^{(8,1)}_2\;\str(\Lambda X_+)\ , \label{pqs} \\
Q^{PQA}_{penguin}&\rightarrow&f^2\;\alpha^{(8,8)}\;
\str(\Lambda\Sigma A\Sigma^\dagger)
\ , \label{pqa}
\ea
where
\be
L_\mu=i\Sigma\partial_\mu\Sigma^\dagger\ ,\ \ \ \ \
X_\pm=2B_0(\Sigma M^\dagger\pm M\Sigma^\dagger)\ ,
\label{bb}
\ee
with $M$ the quark-mass matrix, $B_0$ the parameter $B_0$ of ref.~\cite{gl},
$\Sigma=\Exp(2i\Phi/f)$ the unitary field describing the partially quenched
Goldstone-meson multiplet, and $f$ the bare pion-decay constant
normalized such that $f_\pi=132$~MeV.  The $\alpha$'s are the corresponding
LECs.  Notice that $Q^{PQA}_{penguin}$,
unlike $Q^{PQS}_{penguin}$, is of order $p^0$, due to the fact that
the right-handed current in $Q^{PQA}_{penguin}$
is not a partially quenched singlet (\cf\ electro-magnetic
penguins\footnote{In fact, $Q^{PQA}_{penguin}$ is a component of the same
irrep as the electro-magnetic penguin, except for $N=0$ \cite{gp}.}).  
As already observed in ref.~\cite{gp}, the new operator
$Q^{PQA}_{penguin}$ does not contribute at tree level to matrix elements
with only valence quarks on external lines, since the matrix $A$
is effectively proportional  to the unit matrix 
in the valence sector.  Indeed, 
replacing $A$ by the unit matrix in eq.~(\ref{pqa}) would make the operator
vanish.  This is no longer true at next-to-leading-order (NLO), \ie\ at order 
$p^2$, where one-loop contributions from $Q^{PQA}_{penguin}$ to valence-quark 
matrix elements are non-zero.

Since the singlet-operator contributions also start at order $p^2$,
the NLO contributions from $Q^{PQA}_{penguin}$
compete with the LO contributions from
$Q^{PQS}_{penguin}$, and thus need to be taken into account already
in a leading-order 
analysis of $K\to 0$, $K\to\pi$ and $K\to\pi\pi$ matrix elements.
This also implies that a renormalization scale dependence already appears  
at leading order for those partially quenched matrix elements.
That scale dependence is absorbed by new $O(p^2)$ counterterms for 
$Q^{PQA}_{penguin}$. The complete
list of $CPS$-even \cite{cbetal} operators  is 
\ba
Q^{PQA}_1&=&\frac{\beta^{(8,8)}_1}{(4\pi)^2}\;
\str(\Lambda\{\Sigma A\Sigma^\dagger,L_\mu L_\mu\})\ ,
\label{nlo}\\
Q^{PQA}_2&=&\frac{\beta^{(8,8)}_2}{(4\pi)^2}\;
\str(\Lambda L_\mu\Sigma A\Sigma^\dagger L_\mu)\ ,
\nonumber \\
Q^{PQA}_3&=&\frac{\beta^{(8,8)}_3}{(4\pi)^2}\;
\str(\Lambda\{\Sigma A\Sigma^\dagger,X_+\})\ ,
\nonumber\\
Q^{PQA}_4&=&\frac{\beta^{(8,8)}_4}{(4\pi)^2}\;
\str(\Lambda[\Sigma A\Sigma^\dagger,X_-])\ ,
\nonumber\\
Q^{PQA}_5&=&\frac{\beta^{(8,8)}_5}{(4\pi)^2}\;
\str(\Lambda\Sigma A\Sigma^\dagger)\;\str(L_\mu L_\mu)\ , 
\nonumber\\
Q^{PQA}_6&=&\frac{\beta^{(8,8)}_6}{(4\pi)^2}\;
\str(\Lambda\Sigma A\Sigma^\dagger)\;\str(X_+)\ ,
\nonumber\\
Q^{PQA}_7&=&\frac{\beta^{(8,8)}_7}{(4\pi)^2}\;
i\;\partial_\mu\;\str(\Lambda[\Sigma A\Sigma^\dagger, L_\mu])\ ,
\nonumber
\ea
where we introduced the $O(p^2)$ LECs $\beta^{(8,8)}_{1,\dots,7}$.

The partially quenched theory with $N=3$ light sea quarks
represents a special case.
For $N=3$, the LECs
of the partially quenched theory must be the same as those of the physical,
unquenched theory \cite{shsh}, basically because they represent the
coefficients in an expansion in powers of quark masses, and thus
only depend on the number of dynamical (sea) quarks, and not
on the quark masses themselves.  Therefore, in the $N=3$ partially quenched theory,
what one should do is to omit $Q^{PQA}_{penguin}$ altogether,
because the aim is to obtain the values of $\alpha^{(8,1)}_{1,2}$,
and not the $(8,8)$ LECs $\alpha^{(8,8)}$ and $\beta_i^{(8,8)}$.%
\footnote{As long as one does not consider electro-magnetic penguin 
contributions.}

The quenched case, $N=0$, is different. In this case, the
decomposition reads
\ba
Q^{QCD}_{penguin}&=&
\frac{1}{2}\;\str(\Lambda\psi\psibar\gamma_\mu P_L)
\;\str(\psi\psibar\gamma_\mu P_R)+
\str(\Lambda\psi\psibar\gamma_\mu P_L)
\;\str(\Nh\psi\psibar\gamma_\mu P_R)\ , \nonumber \\
&\equiv&\frac{1}{2}\;Q^{QS}_{penguin}+Q^{QNS}_{penguin}\ , \label{qdecomp}\\
\Nh&=&\frac{1}{2}\diag(1,1,1,
-1,-1,-1)\ , \label{s}
\ea
where valence entries of $\Nh$ are equal to $\frac{1}{2}$,
and ghost entries are equal to $-\frac{1}{2}$.  
The first operator in the decomposition is a singlet under $SU(3|3)_R$, 
while the second is not ($NS$ for non-singlet). 
However, $Q^{QNS}_{penguin}$ can 
transform into the singlet operator, implying that
the non-singlet operators do not form a representation by themselves.
In other words, $Q^{QNS}_{penguin}$ can mix into
$Q^{QS}_{penguin}$, which is possible because $\Nh$ is not supertrace-less,
unlike $A$ in the partially quenched case.  The ChPT realization of both
operators is obtained from the expressions given in eqs.~(\ref{pqa},\ref{nlo})
and by replacing $A\to\Nh$.  When referring to
the quenched theory, we will add a subscript $q$ to the LECs, and
rename\footnote{In
the quenched theory there is no relation between $Q^{QNS}_{penguin}$
and the electro-magnetic penguins \cite{gp}.}
the LECs for $Q^{QNS}_{penguin}$ as $\alpha^{(8,8)}
\to\alpha_q^{NS}$, $\beta_i^{(8,8)}\to\beta_{qi}^{NS}$.

Within the fully quenched approximation, as also in the partially quenched 
case with $N\ne 3$, there is no
reason that the LECs should have the same values
as those of the unquenched theory.  In general, the non-analytic terms
are modified by quenching, and even the scale
dependence of LECs is different between the quenched and unquenched
theories.  It is thus not {\it a priori} clear what choice to
make for the embedding of penguin operators into the quenched theory.
We will return to this issue in section 4 below.  
\section{\large\bf
 Non-singlet \boldmath{$K^0\to\pi^+\pi^-$} matrix elements with general 
kinematics}
\secteq{3}
In this section we present the partially quenched and quenched results
for the contribution of strong penguin operators to the $K^0\to\pi^+\pi^-$
matrix element to order $p^2$ in ChPT.  
We will restrict ourselves to the isospin limit in
the valence sector, $m_{vu}=m_{vd}$, not assuming momentum
conservation, so that $q\ne p_1+p_2$, with $q$ the (ingoing) $K^0$
momentum and $p_1$ ($p_2$) the (outgoing) $\pi^+$ ($\pi^-$) momentum.
All momenta are onshell and we work in euclidean space, \ie\
$p_1^2=p_2^2=-M_\pi^2$, $q^2=-M_K^2$.  
$M_{jvi}$ ($M_{jsi}$) is the mass of a meson made out of the
$j$th valence quark and the $i$th valence (sea) quark.
In all results presented below, we have symmetrized the expressions
in the pion momenta $p_1$ and $p_2$.

\begin{figure}[htb]
\centerline{\mbox{\epsfxsize=13cm\epsfysize=2cm\epsffile{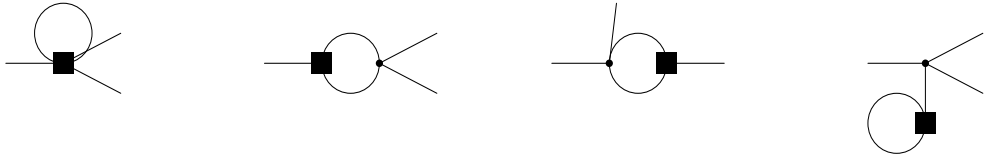}}}
\caption{\small\em
  One-loop diagrams with the insertion of the non-singlet operator
at the weak vertex (box) contributing to  $K^0\to\pi^+\pi^-$ with general 
kinematics}
\label{figure}
\end{figure}

For the contribution of the singlet operator in eq.~(\ref{pqs})
a simple tree-level calculation yields the order $p^2$ result
\ba
\langle\pi^+\pi^-|Q^{PQS}|K^0\rangle &=&
-\frac{4i}{f^3}\Biggl\{
\alpha^{(8,1)}_1\left(\frac{1}{2}q(p_1+p_2)+p_1 p_2\right)
\label{kpps}\\
&& +\frac{2}{3}\alpha^{(8,1)}_2(M_K^2-M_\pi^2)
\left(1+\frac{1}{2}\,\frac{2p_1p_2+q(p_1+p_2)-2M_\pi^2}
{(q-p_1-p_2)^2+M_K^2}\right)\Biggr\}\ .\nonumber
\ea
Note that the contribution proportional to $\alpha^{(8,1)}_2$
vanishes for energy-conserving kinematics, because the corresponding
operator can be written as a total derivative by using equations of
motion \cite{cbetal}.
As mentioned before,
the tree-level contribution from the new non-singlet operator in 
eq.~(\ref{pqa}) vanishes. 
At one loop, it contributes at order $p^2$.  For general kinematics, 
we obtain, in the isospin limit,
\ba
\langle\pi^+\pi^-|Q^{PQA}|K^0\rangle_{1-loop} &=& 
\frac{1}{2}[{\cal A}(q;p_1,p_2)+{\cal A}(q;p_2,p_1)]\ ,\label{kpploop} \\
{\cal A}(q;p_1,p_2)&=&
\frac{4i}{3f^3}\,\alpha^{(8,8)}
\sum_{i~sea}\Biggl\{
3(p_1 p_2 - M_\pi^2)\, I(M_{2si}^2,M_{2si}^2, (p_1+p_2)^2)\nonumber\\
&&
+\frac{2M_K^2-M_\pi^2+qp_1}{(q-p_1)^2}\, \left (
L(M_{2si}^2) - L(M_{3si}^2) \right )\nonumber\\
&&
-6\;\frac{(qp_1+M_K^2)(qp_1+M_\pi^2)}{(q-p_1)^2}\, 
I(M_{2si}^2,M_{3si}^2, (q-p_1)^2)\nonumber\\
&&
+\frac{ (2p_1p_2-2qp_1+4qp_2-2M_\pi^2)}{(q-p_1-p_2)^2+M_K^2}\,
(L(M_{3si}^2) - L(M_{2si}^2) ) \Biggr\}\  \nonumber\\
&&
-\sum_{i~ ghost}(M_{jsi}\to M_{jvi},\, j=2,3)\, , \nonumber
\ea
where we used that $M_{3si}^2-M_{1si}^2=M_K^2-M_\pi^2$ at this order, and
\ba
L(M^2)&=&\int\frac{d^D\ell}{(2\pi)^D}\,\frac{1}{\ell^2+M^2}\ ,\label{integrals}
\\
I(M_1^2, M_2^2,p^2)&=&-\int\frac{d^D\ell}{(2\pi)^D}\,
\frac{1}{(\ell^2+M_1^2)((\ell-p)^2+M_2^2)}\ ,\nonumber
\ea
where $D$ is the number of space-time dimensions.  Explicit 
 expressions for these integrals in the $\overline{MS}$ scheme
are given in the appendix.
The one-loop diagrams contributing to this result are displayed in 
fig.~(\ref{figure}).
In particular, the vacuum-tadpole diagram (last diagram in 
fig.~(\ref{figure})) gives a non-zero 
contribution for general kinematics which corresponds to
the next-to-last line of eq.~(\ref{kpploop}). 
In order to obtain the fully quenched result with no sea quarks at all, one
simply drops the sum over sea quarks in eq.~(\ref{kpploop}), 
keeping only the sum over ghost quarks, and replaces 
$\alpha^{(8,8)}\to\alpha^{NS}_q$.  
There are no contributions from $\eta'$ double-poles.

The new $O(p^2)$ counterterms in eq.~(\ref{nlo}) give,
again for general kinematics
\ba
\langle\pi^+\pi^-|Q^{PQA}|K^0\rangle_{ct}&=&
\left(1-\frac{3}{N}\right)\frac{4i}{(4\pi^2)f^3}\Biggl\{
\left (2\beta^{(8,8)}_1+\beta^{(8,8)}_2\right )\left(\frac{1}{2}
q(p_1+p_2)+p_1p_2\right)
\nonumber\\
&&\hspace{-0.5cm}-\frac{4}{3}\beta^{(8,8)}_3 (M_K^2-M_\pi^2)
\left ( 1 +\frac{1}{2}\,
\frac{2p_1p_2+q(p_1+p_2)-2M_\pi^2}{(q-p_1-p_2)^2+M_K^2}
\right )
\Biggr\}\ .
\label{kpplec}
\ea
The structure of these contributions is such that the scale dependence
contained in eq.~(\ref{kpploop}) can be fully compensated by the LECs
$\beta^{(8,8)}_i$, $i=1,2,3$.  It is easy to verify that the
other LECs $\beta^{(8,8)}_i$, $i=4,\dots,7$ do not contribute.
Since external legs only contain valence quarks, the tree-level
 $K\to\pi\pi$ matrix element does not involve the diagonal
$A_{ii}$ elements of the tensor $A$ with $i$ referring to a sea
or ghost quark.  This implies that, for this calculation, we may
replace $A$ with the unit matrix, and the operators
$Q^{PQA}_i$ in eq.~(\ref{pqs}) for $i=4,\dots,7$ vanish,
while they become proportional
to the operators in eq.~(\ref{pqs}) for $i=1,2,3$.
It follows that the LECs
$\alpha^{(8,1)}_{1,2}$ together with $\beta^{(8,8)}_{1,2,3}$ will 
always appear in the specific combinations 
\ba
&&\alpha^{(8,1)}_1-\left(1-\frac{3}{N}\right)\frac{1}{(4\pi)^2}
(2\beta^{(8,8)}_1+\beta^{(8,8)}_2)\ ,\label{comb}\\
&&\alpha^{(8,1)}_2+\left(1-\frac{3}{N}\right)\frac{2}{(4\pi)^2}
\beta^{(8,8)}_3 \nonumber
\ea
for all tree-level matrix elements with only valence quarks on the
external legs.
For the analogue of eq.~(\ref{comb}) in the quenched case,
 one replaces the factor $1-3/N$ by $1/2$,
 $\alpha^{(8,1)}_i\to\alpha^{(8,1)}_{qi}$, 
$\alpha^{(8,8)}\to\alpha^{NS}_q$ and $\beta^{(8,8)}_i\to\beta^{NS}_{qi}$.

We conclude this section with the expressions for the same
matrix elements in the case of ``physical" kinematics, \ie\ with
the choice $q=p_1+p_2$.  Setting $q=p_1+p_2$ and using that $q^2=-M_K^2$
and $p_1^2=p_2^2=-M_\pi^2$, one obtains from eqs.~(\ref{kpps}),
(\ref{kpploop}) and (\ref{kpplec})
\ba
\langle\pi^+\pi^-|Q^{PQS}|K^0\rangle^{phys} &=& \frac{4i}{f^3}
\alpha^{(8,1)}_1(M_K^2-M_\pi^2)\ ,\label{kppphys1}\\
\langle\pi^+\pi^-|Q^{PQA}|K^0\rangle_{1-loop}^{phys} &=& 
\frac{4i}{3f^3}\,\alpha^{(8,8)}
\sum_{i~sea}\Biggl\{
-\frac{3}{2}M_K^2\, I(M_{2si}^2,M_{2si}^2, -M_K^2)\label{kppphys2}\\
&& -\left(\frac{3}{2}\frac{M_K^4}{M_\pi^2}-3M_K^2\right)
\, I(M_{2si}^2,M_{3si}^2, -M_\pi^2)
\nonumber\\
&& -\left(\frac{3}{2}\frac{M_K^2}{M_\pi^2}-3\right)\, \left [
L(M_{2si}^2) - L(M_{3si}^2) \right ] \Biggr\}\ \nonumber\\
&& -\sum_{i~ ghost}(M_{jsi}\to M_{jvi},\, j=2,3)\ , \nonumber\\
\langle\pi^+\pi^-|Q^{PQA}|K^0\rangle_{ct}^{phys}&=&
\!\!\frac{-4i}{(4\pi^2)f^3}
\left(1\!-\!\frac{3}{N}\right)
\left (2\beta^{(8,8)}_1\!+\!\beta^{(8,8)}_2\right )(M_K^2-M_\pi^2)
\ .
\label{kppphys3}
\ea
Dropping the sea-quark terms and carrying out the sum over ghost quarks,
we obtain a more explicit expression for the one-loop contribution in 
the quenched case, in the $\overline{MS}$ scheme:
\ba
\langle\pi^+\pi^-|Q^{QNS}|K^0\rangle_{1-loop}^{phys}&=&\label{qphys}\\
&&\hspace{-2.7cm}\frac{i}{16\pi^2 f^3}\alpha^{NS}_q\left\{12(M_K^2-M_\pi^2)
\left(\log{\frac{M_\pi^2}{\Lambda^2}}-1\right) 
\right.\nonumber\\
&&\hspace{-0.2cm}
+\left(\frac{M_K^6}{M_\pi^4}-2\frac{M_K^4}{M_\pi^2}+2M_K^2\right)
\log{\frac{M_K^2}{M_\pi^2}}\nonumber\\
&&\hspace{-0.2cm}
+\left(\frac{M_K^6}{M_\pi^4}-6\frac{M_K^4}{M_\pi^2}+10M_K^2-4M_\pi^2\right)
\log{\frac{2M_K^2-M_\pi^2}{M_\pi^2}}\nonumber\\
&&\hspace{-2.7cm}
+2M_K^2 \left(F(M_\pi^2,M_\pi^2,-M_K^2)-2i\,\pi\,\theta(M_K^2-4M_\pi^2)\,
\sqrt{1-\frac{4M_\pi^2}{M_K^2}}
+{\frac{\pi}{3}\sqrt{3}}\right)
\nonumber\\
&&\hspace{-2.7cm}
\left.
+\left(\frac{M_K^4}{M_\pi^2}-2M_K^2\right)(2F(M_\pi^2,M_K^2,-M_\pi^2)
+F(M_K^2, 2M_K^2-M_\pi^2,-M_\pi^2))\right\}\nonumber\ ,
\ea
where the function $F$ is given in the appendix.
\section{\large\bf Strategies for quenched estimates of real-world penguin
matrix elements}
\secteq{4}
Recent numerical estimates of $K\to\pi\pi$ matrix elements
reported in refs.~\cite{cppacs,rbc} have been obtained via the {\em indirect}
method, where the simpler $K\to\pi$ and $K\to 0$ transition amplitudes are 
computed on the lattice and then converted into estimates for $K\to\pi\pi$
matrix elements using ChPT.
However, the fact that those numerical results are still obtained in the 
quenched approximation introduces a source of systematic error which is in 
principle uncontrolled.  
As already explained, the LECs of the quenched theory do
not have to have values equal to those of the unquenched theory.
Typically, even the scale dependence of quenched and unquenched
LECs is not the same; it depends on the number of light dynamical (sea) quarks
in the theory.  

In the case of penguin operators, an additional
ambiguity arises because it is {\it a priori} unclear whether it
would be best to take matrix elements of $Q^{QCD}_{penguin}$, \ie\ a linear 
combination of $Q^{QS}_{penguin}$ and $Q^{QNS}_{penguin}$ as in 
eq.~(\ref{qdecomp}), or to drop the
contribution from $Q^{QNS}$ under the assumption that 
$\alpha^{(8,1)}_{q1}$ is the best estimate of $\alpha^{(8,1)}_1$.

In order to discuss possible strategies in more detail, we
first recall the leading order ChPT expressions for $K\to\pi$
and $K\to 0$ matrix elements of strong penguin operators, from 
ref.~\cite{gp}. In the quenched approximation ($N=0$) one has
\ba
\langle\pi^+|Q^{QCD}_{penguin}|K^+\rangle
&\!\!=\!\!&\frac{4M^2}{f^2}\left\{
\alpha^{(8,1)}_{q1}-\alpha^{(8,1)}_{q2}-\frac{1}{(4\pi)^2}
\left(\beta^{NS}_{q1}+\frac{1}{2}\beta^{NS}_{q2}+\beta^{NS}_{q3}
\right)
\right\}\,,\label{ktopi}\\
\langle 0|Q^{QCD}_{penguin}|K^0\rangle
&\!\!=\!\!&\frac{4i}{f}\left\{\left(
\alpha^{(8,1)}_{q2}+\frac{1}{(4\pi)^2}\beta^{NS}_{q3}\right)
(M_K^2-M_\pi^2)\right.\label{ktovac}\\
&&\left.\ \ \ \ \ \ \  
+\alpha^{NS}_q\sum_{i\ valence}\left(L(M_{3vi}^2)-L(M_{2vi}^2)
\right)\right\}\ ,\nonumber
\ea
where $M_K=M_\pi=M$ in the case of the $K\to\pi$ matrix element,
and contributions of both singlet and non-singlet operators
are included.  Notice that only the (quenched versions of the)
combinations (\ref{comb}) of LECs appear in these expressions, 
as expected.
Assuming that one can limit the analysis to leading-order 
in ChPT, there are at least three different strategies for estimating
$K\to\pi\pi$ penguin matrix elements from LECs obtained by
fitting eqs.~(\ref{ktopi},\ref{ktovac}) to quenched numerical
results: 

\begin{itemize}
\item[1.]  Ignore $\alpha^{NS}_q$, but not the other LECs associated
with the non-singlet operator, $\beta^{NS}_{q1,2,3}$.  Both
$\beta^{NS}_{q3}$ and $\beta^{NS}_{q1}+\frac{1}{2}\beta^{NS}_{q2}$ are scale
dependent (however, their sum is not, as can be seen from 
eq.~(\ref{ktopi})), implying that this strategy is scale dependent.  
However, it still makes sense in case the non-analytic contribution
 proportional to $\alpha^{NS}_q$ 
(\cf\ eq.~(\ref{kppphys2})) is numerically small compared to
all other contributions
at a reasonable scale $\Lambda$ of order 1~GeV.  Thus, the linear combination
$\alpha^{(8,1)}_{q1}-\frac{1}{(4\pi)^2}(\beta^{NS}_{q1}
+\frac{1}{2}\beta^{NS}_{q2})$ is taken as the best estimate
for the unquenched $\alpha^{(8,1)}_1$, and eq.~(\ref{kppphys1}) 
can then be used to obtain the physical $K\to\pi\pi$ matrix element 
(at tree level).
This is the strategy followed in refs.~\cite{cppacs,rbc}.
In fact, in these works, it was assumed 
that the contribution proportional to $\alpha^{NS}_q$ in 
eq.~(\ref{ktovac}) is small. 
\item[2.] Drop all the non-singlet operators.  It was shown in ref.~\cite{gp}
that this can be done by dropping, in the fully quenched case,
all {\em eye}-diagrams
in which the right-handed quarks in eq.~(\ref{penguin}) are contracted.
This can be easily deduced from eq.~(\ref{qdecomp}).  This strategy
was explored for $Q_6$ in ref.~\cite{lanl}.  (For the partially-quenched
case, see below.)
\item[3.] Perform a complete quenched calculation including all contributions from 
singlet and non-singlet operators. After extracting 
all the LECs, singlet and non-singlet, one can use the sum of 
eqs.~(\ref{kppphys1}), (\ref{qphys}) and the quenched version of 
eq.~(\ref{kppphys3}) 
to determine the quenched $K\to\pi\pi$ matrix element at the physical point.
\end{itemize}

Strategy 2 isolates $\alpha^{(8,1)}_{q1}$, and might thus appear
to be the obvious choice, since it is this LEC that is needed for
calculating $K\to\pi\pi$ matrix element (to chiral leading-order)
in the unquenched theory.
However, as we already mentioned, the values of LECs in the
quenched and unquenched theories do not have to be equal, and
it might happen that (at some scale $\Lambda$) the quenched 
combination $\alpha^{(8,1)}_{q1}-\frac{1}{(4\pi)^2}(\beta^{NS}_{q1}
+\frac{1}{2}\beta^{NS}_{q2})$, determined from strategy 1,
is indeed a better estimate of $\alpha^{(8,1)}_1$.  Strategy 2 can
be viewed as the situation in which the strong interactions are
quenched at all scales between the weak and hadronic scales,
because in that case only singlet penguin operators would
appear in the evolution from the weak to the hadronic scale.
So, while, on the one hand, it appears natural to assume only a mild flavor
dependence of the LECs, in particular $\alpha^{(8,1)}_1$,
one might, on the other hand, 
argue that it is better to calculate the evolution 
from the weak to the hadronic scale in the unquenched theory,
even if the matrix element at the hadronic scale 
is finally computed in the quenched
approximation.  The key point is that it is impossible to
decide which strategy is best.

The exception to these observations is the case
of partially quenched QCD in which the number of
light sea quarks is {\em  equal} to that of the real world, in which
$N=3$.  In the partially quenched theory, the singlet operator is
(\cf\ eq.~(\ref{pqdecomp}))
\be
Q^{PQS}_{penguin}=\frac{3}{N}(\sbar d)_L
\left(\ubar_v u_v+\dbar_v d_v+\sbar_v s_v+
\sum_i \qbar_{si}q_{si}+\ubar_g u_g+\dbar_g d_g+\sbar_g s_g\right)_R\,,
\label{pqsagain}
\ee
where the subscripts $v$, $s$ and $g$ denote valence, sea and ghost quarks,
respectively.  Strategy 2 now corresponds to dropping all diagrams
in which the right-handed valence and ghost quarks in the second
factor of eq.~(\ref{pqsagain}) are contracted
\cite{gp}.  If the number of sea quarks $N=3$ (but with the sea- and
valence-quark masses not necessarily equal), the singlet
LECs $\alpha^{(8,1)}_{1,2}$ are those of the real world \cite{shsh},
and therefore strategy 2 is the only correct one in this case.

For any other case, fully quenched or partially quenched with $N\neq 3$,
there is {\it a priori} no preferred choice; the spread in results obtained 
by employing all three strategies should be taken as a (lower bound of the) 
systematic error due to quenching.
The extent to which strategies 1 and 3 lead to numerically different results 
depends
on the size of $\alpha^{NS}_q$ contributions (at a given scale $\Lambda$).
{}From eqs.~(\ref{kppphys1},\ref{qphys}) we find, taking physical 
values for all parameters, $M_K=500$~MeV, $M_\pi=140$~MeV,
$f=f_\pi=132$~MeV, $M_\rho=770$~MeV and $M_\eta=550$~MeV, that
\ba
-i[K^0\to\pi^+\pi^-]_q&=&
400.7(\alpha^{(8,1)}_{q1}-\alpha^{(27,1)}_q)+(28.2-7.2i)\alpha^{NS}_q
\ \ \ (\Lambda=1\ {\rm GeV})\ ,\nonumber\\
&=&400.7(\alpha^{(8,1)}_{q1}-\alpha^{(27,1)}_q)+(32.2-7.2i)\alpha^{NS}_q
\ \ \ (\Lambda=M_\rho)\ ,\nonumber\\
&=&400.7(\alpha^{(8,1)}_{q1}-\alpha^{(27,1)}_q)+(37.3-7.2i)\alpha^{NS}_q
\ \ \ (\Lambda=M_\eta)\ ,
\label{qnum}
\ea
where we added in the tree-level ChPT contribution from the 
$SU(3)_L$ 27-plet operator \cite{cbetal}.  If $\alpha^{NS}_q$ is
of the same order as $\alpha^{(8,1)}_{q1}-\alpha^{(27,1)}_q$, the
contribution of the terms proportional to $\alpha^{NS}_q$ is indeed
small.  The smallness of the coefficient of $\alpha^{NS}_q$ is due
to a $1/(4\pi)^2$ suppression factor coming from the loop integral,
and one might argue that $\alpha^{NS}_q/(4\pi)^2$ is the 
``natural" parameter to compare with $\alpha^{(8,1)}_{q1}-\alpha^{(27,1)}_q$,
in which case the contribution would not be small.  
Notice also that a small spurious imaginary part is generated by the 
non-singlet operator via the ghost-pion one-loop rescattering diagram.
It is clear
that the value of $\alpha^{NS}_q$ will have to be determined from
a lattice computation.  While this can be done by including the
$\alpha^{NS}_q$ terms, of {\em e.g.} eq.~(\ref{ktovac}), in a fit
to lattice data, there exists a much simpler and more reliable
way of estimating the size of $\alpha^{NS}_q$, as will be
explained in the next section.  

Under the assumption that  $\alpha^{NS}_q$ can be neglected without 
 introducing a large uncertainty into the final estimate of
strong penguin $K\to\pi\pi$ matrix elements, the question 
remains whether (to leading order in ChPT)
$\alpha^{(8,1)}_{q1}$ or $\alpha^{(8,1)}_{q1}-\frac{1}{(4\pi)^2}
(\beta^{NS}_{q1}+\frac{1}{2}\beta^{NS}_{q2})$ would be a better
estimate of $\alpha^{(8,1)}_1$.
 The issue was investigated in ref.~\cite{lanl},
where it was found that the difference between the two choices
is numerically significant.
At the physical kaon mass, the numerical value
of the $B$ parameter corresponding to $Q_6$ turns out to be
 approximately twice
as large when the contribution of the non-singlet operator
$Q^{QNS}_{penguin}$ is omitted altogether.  Translated into estimates
for the leading-order LECs, this implies that $\alpha^{(8,1)}_{q1}$
is approximately twice as large as $\alpha^{(8,1)}_{q1}-\frac{1}{(4\pi)^2}
(\beta^{NS}_{q1}+\frac{1}{2}\beta^{NS}_{q2})$.  This may lead to 
substantial modifications in quenched estimates of $\eps'/\eps$,
as discussed in ref.~\cite{gplat2002}.  

We emphasize that the whole
discussion here is based on leading-order ChPT, and that NLO
contributions may still lead to a substantial correction.  However, it is 
reasonable to  believe that NLO effects will not 
invalidate the basic content of our observations.

Finally, we give a few more numerical examples of
the partially quenched case with $N=2$, always keeping the
valence quark masses at their physical values (in the isospin limit), 
and choosing the two sea quarks to be degenerate in mass.  Taking
$m_{sea}=m_u=m_d$, we find
\ba
-i[K^0\to\pi^+\pi^-]_{N=2}&=&
400.7(\alpha^{(8,1)}_1-\alpha^{(27,1)})+12.7\alpha^{(8,8)}
\ \ \ (\Lambda=1\ {\rm GeV})\ ,\nonumber\\
&=&400.7(\alpha^{(8,1)}_1-\alpha^{(27,1)})+14.0\alpha^{(8,8)}
\ \ \ (\Lambda=M_\rho)\ ,\nonumber\\
&=&400.7(\alpha^{(8,1)}_1-\alpha^{(27,1)})+15.7\alpha^{(8,8)}
\ \ \ (\Lambda=M_\eta)\ ,
\label{qnumpq}
\ea
whereas taking $m_{sea}=m_s$ we obtain
\ba
-i[K^0\to\pi^+\pi^-]_{N=2}&=&
400.7(\alpha^{(8,1)}_1-\alpha^{(27,1)})+(2.8-7.2i)\alpha^{(8,8)}
\ \ \ (\Lambda=1\ {\rm GeV})\ ,\nonumber\\
&=&400.7(\alpha^{(8,1)}_1-\alpha^{(27,1)})+(4.1-7.2i)\alpha^{(8,8)}
\ \ \ (\Lambda=M_\rho)\ ,\nonumber\\
&=&400.7(\alpha^{(8,1)}_1-\alpha^{(27,1)})+(5.8-7.2i)\alpha^{(8,8)}
\ \ \ (\Lambda=M_\eta)\ .
\label{qnumpqs}
\ea
Recall that for the $N=2$ theory values of the LECs do not
have to equal those of the $N=3$ theory. However, the partially quenched
theory with two light sea quarks 
is closer to the real-world theory than the quenched ($N=0$) theory. 
This is reflected by the fact that the coefficients of $\alpha^{(8,8)}$
are small compared to those of $\alpha_q^{NS}$ in eq.~(\ref{qnum}).  
Notice in addition that in the $N=2$ case, with $m_{sea}=m_u=m_d$, 
the small spurious imaginary part
vanishes, since it comes entirely from the pion-rescattering loop
diagram where the sea-quark contribution is now fully cancelled by the 
corresponding ghost-quark contribution.

In the case of $N=3$ sea quarks with
masses equal to the three valence quarks,  ghost- and sea-quark
contributions in eqs.~(\ref{kpploop},\ref{kpplec})
cancel,\footnote{In eq.~(\ref{kpplec}) the cancellation already
occurs by just setting $N=3$, because in this tree-level expression
the sea- and ghost-quark masses do not appear.} \P
as they should, because this choice of parameters corresponds
precisely to unquenched QCD.  

\section{\large\bf How to determine $\alpha^{NS}_q$ on the lattice}
\secteq{5}
In principle, it is possible to determine $\alpha^{NS}_q$ from matrix
elements with only physical (valence) particles as external states.  
For instance, given good enough statistics and a wide enough
range of quark masses, it can be determined
from a fit to eq.~(\ref{ktovac}).  However, as also pointed out in
ref.~\cite{rbc}, the logarithmic terms in eq.~(\ref{ktovac}) can 
look very linear in the typical range of quark masses used in
lattice computations, making it hard to disentangle $\alpha^{NS}_q$
from $\alpha^{(8,1)}_{q2}+\frac{1}{(4\pi)^2}\beta^{(8,1)}_{q3}$.  
It would therefore be preferable to 
determine $\alpha^{NS}_q$ from a matrix element to which it
contributes at order $p^0$, because no other operators can
``contaminate" the result at that order.

It is very simple to do so, by considering matrix elements with
ghost quarks on the external lines instead of valence quarks.  
Since this corresponds to a flavor
rotation on the external lines, one needs to rotate the
operator $Q^{QNS}_{penguin}$ accordingly.  A key point is that,
while of course ghost quarks are not explicitly present in a
quenched computation, their propagators are identical to 
those of the valence quarks, which are available in the actual
computation.

So, in order to determine $\alpha^{NS}_q$, we propose to consider
the following matrix element.  First, we rotate $Q^{QNS}_{penguin}$
by an $SU(3|3)_L$ rotation into
\be
{\tilde Q}^{QNS}_{penguin}=
(\sbar\gamma_\mu P_L\td)\,\str(\Nh\psi\psibar\gamma_\mu P_R) \ . \label{qnsrot}
\ee
This operator is in the same irrep of the group $SU(3|3)_L\times SU(3|3)_R$,
is thus parametrized by the same LECs as $Q^{QNS}_{penguin}$,
and in particular, to leading order, by $\alpha^{NS}_q$.
We then consider the matrix element of this operator between a
{\it fermionic} kaon $\tilde K\propto{\tdb}\gamma_5 s$
and the vacuum.  To leading order,
\be
\langle 0|{\tilde Q}^{QNS}_{penguin}|\tilde K\rangle=
2if\alpha^{NS}_q+O(p^2) \ ,\label{alphans}
\ee
thus isolating $\alpha^{NS}_q$.  Carrying out all quark Wick contractions,
one finds that
\ba
\langle 0|{\tilde Q}^{QNS}_{penguin}(y){\tdb}(x)\gamma_5 s(x)
|0\rangle&=& \label{contraction}\\
&&\hspace{-5.6cm}-\frac{1}{2}\left\{
\tr\Bigl[\gamma_5\langle s(x)\sbar(y)\rangle\gamma_\mu P_L
\langle\td(y)\tdb(x)\rangle\Bigr]\tr\Bigl[
\gamma_\mu P_R(\langle u(y)\ubar(y)\rangle
+\langle d(y)\dbar(y)\rangle+\langle s(y)\sbar(y)\rangle
\right.\nonumber\\
&&\hspace{2cm}
+\langle \tu(y)\tub(y)\rangle+\langle \td(y)\tdb(y)\rangle
+\langle \ts(y)\tsb(y)\rangle)\Bigr] \nonumber\\
&&-\tr\Bigl[\gamma_5\langle s(x)\sbar(y)\rangle\gamma_\mu P_R
\langle s(y)\sbar(y)\rangle\gamma_\mu P_L\langle\td(y)\tdb(x)\rangle\Bigr]
\nonumber\\
&&\left.+\tr\Bigl[\gamma_5\langle s(x)\sbar(y)\rangle\gamma_\mu P_L
\langle\td(y)\tdb(y)\rangle\gamma_\mu P_R
\langle\td(y)\tdb(x)\rangle\Bigr]\right\} \ ,\nonumber
\ea
where the traces are over spin and color indices only.  A key
observation is now 
that ghost propagators and valence propagators are equal 
flavor by flavor, $\langle\td(y)\tdb(x)\rangle=\langle d(y)\dbar(x)\rangle$
, \etc. Using this property, eq.~(\ref{contraction}) simplifies to
\ba
\langle 0|{\tilde Q}^{QNS}_{penguin}(y){\tdb}(x)\gamma_5 s(x)
|0\rangle&=& \label{contractionfinal}\\
&&\hspace{-5.4cm}
-\tr\left[\gamma_5\langle s(x)\sbar(y)\rangle\gamma_\mu P_L
\langle d(y)\dbar(x)\rangle\right]\tr\left[
\gamma_\mu P_R(\langle u(y)\ubar(y)\rangle
+\langle d(y)\dbar(y)\rangle+\langle s(y)\sbar(y)\rangle)\right]
\nonumber\\
&&\hspace{-0.1cm}
+\frac{1}{2}\tr\left[\gamma_5\langle s(x)\sbar(y)\rangle\gamma_\mu P_R
\langle s(y)\sbar(y)\rangle\gamma_\mu P_L\langle d(y)\dbar(x)\rangle\right]
\nonumber\\
&&\hspace{-0.1cm}
-\frac{1}{2}\tr\left[\gamma_5\langle s(x)\sbar(y)\rangle\gamma_\mu P_L
\langle d(y)\dbar(y)\rangle\gamma_\mu P_R
\langle d(y)\dbar(x)\rangle\right] \ .\nonumber
\ea
We conclude that it is possible to estimate $\alpha^{NS}_q$ 
as a leading-order effect using only combinations of contractions
of valence-quark propagators.  For the $K\to 0$ matrix element of
$Q^{QNS}_{penguin}$, \cf\ eqs.~(\ref{qdecomp},\ref{ktovac}), 
the contractions in
terms of valence quarks are of the same form, but the first two terms
have the opposite sign, while the
last term has the same sign as in eq.~(\ref{contractionfinal}).
Since the $K\to 0$ matrix
element is of order $p^2$, we may combine the two results to
obtain
\ba
-\tr\left[\gamma_5\langle s(x)\sbar(y)\rangle\gamma_\mu P_L
\langle d(y)\dbar(y)\rangle\gamma_\mu P_R
\langle d(y)\dbar(x)\rangle\right]_{amputated}\label{trick} \\
&&\hspace{-3.1cm}=
\sqrt{Z}\left(\langle 0|{\tilde Q}^{QNS}_{penguin}|\tilde K\rangle
+\langle 0|Q^{QNS}_{penguin}|K\rangle\right) \nonumber \\
&&\hspace{-3.1cm}
=\sqrt{Z}\left(2if\alpha^{NS}_q+O(p^2)\right)\ ,\nonumber
\ea
making it even easier to determine $\alpha^{NS}_q$.
The wave-function renormalization $Z$ is defined from
$\dbar\gamma_5 s =\sqrt{Z}\, K$.  

The analysis
for a similar determination of $\alpha^{(8,8)}$ in the partially
quenched theory is analogous.  There, of course, the observation is
not new, since $\alpha^{(8,8)}$ is also the leading LEC for the 
electro-magnetic penguin (which in the partially quenched theory
with $N\ge 1$ is in the same irrep as $Q^{PQA}_{penguin}$ 
of eq.~(\ref{pqa}) \cite{gp}).
The main differences between a determination of $\alpha^{(8,8)}$ and 
 $\alpha^{NS}_q$ are that, first, $\alpha^{NS}_q$ is {\it not}
related to the electro-magnetic penguin in the quenched case
\cite{gp}, and second, that in order to determine it using
leading-order (in this case $O(p^0)$) 
ChPT, one is forced to consider ghost quarks, as we 
did above.  
\section{\large\bf Conclusion}
\secteq{6}
In this paper, we continued our investigation of the ambiguities
afflicting strong penguin contributions to $K\to\pi\pi$
weak matrix elements due to the use of the quenched approximation.

The fact that the way of {\em embedding} penguin operators of the 
effective weak hamiltonian in the quenched theory is not unique 
tells us that, in the enlarged context of electro-weak interactions,
the usual definition of the quenched theory is not complete.
If only strong interactions are considered, it is sufficient
to define quenched QCD as the modified version of QCD in which 
the quark determinant is set equal to a constant.  A field-theoretic
definition can be given through the introduction of ghost quarks
into the path-integral
\cite{morel}, giving access to a complete picture of the symmetries
of the quenched theory \cite{bgq}.  As soon as one considers
operators external to QCD (\ie\ the addition of electro-weak interactions), 
one has to answer the question how these operators
should be incorporated into the quenched theory.  Usually, this
is straightforward.  One classifies the operator by its
flavor quantum numbers, in other words,
one determines the irrep of $SU(3)_L\times SU(3)_R$ under which 
this operator transforms.  If there exists a larger irrep of the
quenched symmetry group which reduces to the unquenched irrep,
the corresponding component of the quenched irrep can be taken as
the quenched definition of the operator.  However, in the case
of strong penguins the operator, while irreducible in the unquenched
theory, is a linear combination of 
components of {\it more than one} irrep of the quenched symmetry group.
Therefore each LEC of the unquenched theory corresponds to
a set of LECs in the quenched theory.  The ambiguity 
arises, because there is {\it a priori} no criterium for
 which linear combination of quenched LECs (if any) would yield the best 
estimate of the unquenched LEC.   In the case of $LR$ penguins
considered here and in ref.~\cite{gp}, this phenomenon produces an 
effect already at leading order in ChPT.

We remark that even in the simplest case when there exists a one-to-one
correspondence between unquenched and quenched irreps, there is 
still the freedom to choose any component of the quenched irrep,
and this flexibility can be used to extract LECs in the most
convenient way \cite{cs}.  However, in this case there is no
ambiguity in the relation between unquenched and quenched LECs
(even though their values may differ).  This is in principle not
different from the situation within the unquenched theory, where
in general
any component of an irrep can be used to extract the corresponding
LEC.  A classic example  for weak matrix elements 
is the relation between $B_K$ and the
$K^+\to\pi^+\pi^0$ decay rate \cite{dgh}.  (At non-leading order,
it may not be possible to determine all LECs describing an
operator in ChPT from one process, of course.)

The ambiguity affecting penguin operators is fundamental, since
there exists no solid theoretical argument that can be used to
decide the issue.  Therefore, we argue that one should compare
all choices that can be reasonably made, and take the resulting
spread of estimated values as a lower bound on the systematic error due to 
quenching.
It appears that in the case of $\eps'/\eps$ this
systematic error is rather large \cite{lanl,gplat2002}. A leading-order
analysis of currently available lattice data \cite{cppacs,rbc,lanl}
seems to indicate  that quenched lattice computations
cannot even confirm that this parameter is non-vanishing in the
Standard Model.  It could also be that the large numerical
difference found between $\alpha^{(8,1)}_{q1}$ and 
$\alpha^{(8,1)}_{q1}-\frac{1}{(4\pi)^2}(\beta^{NS}_{q1}
+\frac{1}{2}\beta^{NS}_{q2})$ would be 
explained by the fact that higher
orders in ChPT have not been taken into account, but we consider
this to be unlikely.  While it is clear that higher orders
are numerically important, there appears to be no reason to
assume that $\beta^{NS}_{q1}
+\frac{1}{2}\beta^{NS}_{q2}$ is small.  It could also be that
$\alpha_q^{NS}$, which appears in $K^0\to 0$, and needs to be
subtracted to obtain 
$\alpha^{(8,1)}_{q1}-\frac{1}{(4\pi)^2}(\beta^{NS}_{q1}
+\frac{1}{2}\beta^{NS}_{q2})$ from $K^+\to\pi^+$, is not small.
This would affect the determination of $\alpha^{(8,1)}_{q2}
+\frac{1}{(4\pi)^2}\beta^{NS}_{q3}$, and hence the size of the
subtraction.  It is therefore important to obtain a reliable
estimate of $\alpha^{NS}_q$.  We suggested a simple
method for extracting its value.

The above argument does {\it not} imply that lattice 
computations of $\eps'/\eps$ are doomed to fail.
On the contrary, quenched estimates of $\eps'/\eps$
with a particular choice for the strong penguins demonstrate
that this computation is feasible, thanks to major
advances in both theory and computational power.  However,
what will be needed in order to eliminate systematic errors
due to quenching is a partially quenched study with $N=3$
light sea quarks.  This is the only approximation to 
unquenched QCD which is reliable in that it can be
extrapolated systematically to the real world \cite{shsh}.
Currently existing quenched results give us invaluable
information on what is needed to promote them to the required
$N=3$ world.
For partially quenched QCD with $N\ne 3$, the situation is
essentially the same as for quenched QCD, modulo 
differences in detail.

Summarizing, we presented the quenched and partially quenched results
for the non-singlet contribution to the $K^0\to\pi^+\pi^-$ matrix element.
This made it possible to discuss in detail various strategies one
might follow to use quenched computations in order to estimate
the real-world value of this amplitude.  Since
$\alpha^{NS}_q$ contributes to this matrix element, but not to 
the $K^+\to\pi^+$ transition amplitude used in ref.~\cite{cppacs,rbc},
this introduces an additional ambiguity already at leading chiral order.  
The importance of this ambiguity depends on the size of $\alpha^{NS}_q$ and 
we have proposed a simple recipe for its determination.
Our expressions for $K^0\to\pi^+\pi^-$ with the 
most general possible kinematics and the inclusion of the non-singlet 
contributions, are appropriate for the analysis of {\em direct} quenched 
computations of this matrix element at leading order in ChPT. 
Beyond leading order, new problems arise \cite{scalar}, which may invalidate 
current methods for the {\em direct} determination of $K\to\pi\pi$ amplitudes 
with $\Delta I=1/2$ in quenched and partially quenched QCD.
\section*{\large\bf Appendix}
\secteq{A}
In this appendix we collect explicit expressions for the basic loop
integrals appearing in eq.~(\ref{kpploop}) {\it etc.}  
Using dimensional regularization, we have
\be
L(M^2)=\int\frac{d^D\ell}{(2\pi)^D}\,\frac{1}{\ell^2+M^2}
=\frac{M^2}{16\pi^2}\left(
\left[-\frac{2}{\epsilon}+\gamma-\log{4\pi}\right]
+\log\frac{M^2}{\Lambda^2}-1\right)\ ,
\label{lint}
\ee
where $\Lambda$ is the running scale, $\epsilon=4-D$, and 
\ba
\Real\,I(M_1^2, M_2^2,p^2)&=&-\Real\,\int\frac{d^D\ell}{(2\pi)^D}\,
\frac{1}{(\ell^2+M_1^2)((\ell-p)^2+M_2^2)}\label{rei}\\
&&\hspace{-2.5cm}=\frac{1}{16\pi^2}\Biggl\{
\left[-\frac{2}{\epsilon}+\gamma-\log{4\pi}\right]\nonumber\\
&&\hspace{-2.0cm}
-1+\log{\frac{M_2^2}{\Lambda^2}}+\frac{1}{2}
\left(1-\frac{M_1^2}{p^2}+\frac{M_2^2}{p^2}\right)
\log{\frac{M_1^2}{M_2^2}}
+\frac{1}{2}F(M_1^2,M_2^2,p^2)\Biggr\}\ ,\nonumber 
\ea
in which 
\ba
F(M_1^2,M_2^2,p^2)&\!=&\!\sqrt{\lambda\left(1,\frac{M_1^2}{p^2},
\frac{M_2^2}{p^2}\right)}\log{\frac{p^2+M_1^2+M_2^2+p^2\sqrt{\lambda
(1,M_1^2/p^2,M_2^2/p^2)}}%
{p^2+M_1^2+M_2^2-p^2\sqrt{\lambda(1,M_1^2/p^2,M_2^2/p^2)}}}\ ,\nonumber\\
\lambda(x,y,z)&\!=&\!(x-y+z)^2+4xy \ .\label{defs}
\ea
$\overline{MS}$ expressions are obtained by dropping the contact
terms in square brackets.  

For $p^2>0$, the argument of the logarithm in eq.~(\ref{defs})
is positive, and $I$ is real.  For $p^2\le 0$, $F$ is obtained by analytic 
continuation.  $\lambda(1,M_1^2/p^2,M_2^2/p^2)$ turns negative for
$-(M_1+M_2)^2<p^2<-(M_1-M_2)^2$, and we find that $I$ is still real
with $F$ now given by
\be
F(M_1^2,M_2^2,p^2)=2\sqrt{-\lambda\left(1,\frac{M_1^2}{p^2},
\frac{M_2^2}{p^2}\right)}
\arctan{\frac{-p^2\sqrt{-\lambda(1,M_1^2/p^2,M_2^2/p^2)}}{p^2+M_1^2+M_2^2}}
\ .\label{arctan}
\ee
At $p^2=-(M_1^2+M_2^2)$ the argument of the arctangent has a 
singularity, across which the branch of the arctangent has to be
chosen continuously: 
\ba
\arctan{\frac{-p^2\sqrt{-\lambda(1,M_1^2/p^2,M_2^2/p^2)}}{p^2+M_1^2+M_2^2}}
&=&{\rm Arctan}\;{\frac{-p^2\sqrt{-\lambda(1,M_1^2/p^2,M_2^2/p^2)}}
{p^2+M_1^2+M_2^2}
}+\pi\ ,\nonumber \\
&&\hspace{2cm} p^2<-(M_1^2+M_2^2)\ ,\label{cont}
\ea
where Arctan denotes the principal value of the arctangent.
Again continuing analytically across $p^2=-(M_1+M_2)^2$, 
$F$ is again given by eq.~(\ref{defs}), but $I(M_1^2,M_2^2,p^2)$
picks up an imaginary part:
\be
\Imag\,I(M_1^2, M_2^2,p^2)=-\frac{1}{16\pi^2}\,\pi\,
\sqrt{\lambda\left(1,\frac{M_1^2}{p^2},
\frac{M_2^2}{p^2}\right)}\ .\label{imi}
\ee

\section*{\large\bf Acknowledgements}
We would like to thank Norman Christ, Bob Mawhinney, Martin Savage and
Steve Sharpe for discussions.  MG thanks the Institute
for Nuclear Theory at the University of Washington for hospitality,
and EP thanks the Department of Physics and Astronomy at San Francisco
State University for hospitality.  MG is supported in part by the
US Department of Energy, and EP by the Italian MURST under
the program \textit{Fenomenologia delle Interazioni Fondamentali}.


\begin{thebibliography}{99}

\bibitem{cppacs}
J.~Noaki {\it et al.}  [CP-PACS Collaboration],
arXiv:hep-lat/0108013.

\bibitem{rbc}
T.~Blum {\it et al.}  [RBC Collaboration],
arXiv:hep-lat/0110075.

\bibitem{gp}
M.~Golterman and E.~Pallante,
JHEP {\bf 0110}, 037 (2001)
[arXiv:hep-lat/0108010].

\bibitem{cbetal}
C.~Bernard, T.~Draper, A.~Soni, H.~Politzer and M.~Wise,
Phys.\ Rev.\ D {\bf 32}, 2343 (1985).

\bibitem{gplat2002}
M.~Golterman and E.~Pallante,
arXiv:hep-lat/0208069.

\bibitem{bgpq}
C.~Bernard and M.~Golterman,
Phys.\ Rev.\ D {\bf 49}, 486 (1994)
[arXiv:hep-lat/9306005].

\bibitem{replica}
P.~H.~Damgaard and K.~Splittorff, 
Phys.\ Rev.\ D {\bf 62}, 054509 (2000)
[arXiv:hep-lat/0003017].

\bibitem{morel}
A.~Morel,
J.\ Phys.\ (France) {\bf 48}, 1111 (1987).

\bibitem{bgq}
C.~Bernard and M.~Golterman,
Phys.\ Rev.\ D {\bf 46}, 853 (1992)
[arXiv:hep-lat/9204007].

\bibitem{gpchpt}
M.~Golterman and E.~Pallante,
JHEP {\bf 0008}, 023 (2000)
[arXiv:hep-lat/0006029].

\bibitem{gl}
J.~Gasser and H.~Leutwyler,
Nucl.\ Phys.\ B {\bf 250}, 465 (1985).

\bibitem{shsh}
S.~Sharpe and N.~Shoresh,
Phys.\ Rev.\ D {\bf 62}, 094503 (2000)
[hep-lat/0006017].

\bibitem{lanl}
T.~Bhattacharya {\it et al.},
Nucl.\ Phys.\ Proc.\ Suppl.\  {\bf 106}, 311 (2002)
[arXiv:hep-lat/0111004].

\bibitem{cs}
J.~W.~Chen and M.~J.~Savage,
Phys.\ Rev.\ D {\bf 65}, 094001 (2002)
[arXiv:hep-lat/0111050].

\bibitem{dgh}
J.~F.~Donoghue, E.~Golowich and B.~R.~Holstein,
Phys.\ Lett.\ B {\bf 119}, 412 (1982).

\bibitem{scalar}
M.~Golterman and E.~Pallante, Nucl. Phys. Proc. Suppl. 83 (2000) 250, 
[arXiv:hep-lat/9909069];\\
C.~W.~Bernard and M.~F.~Golterman,
Phys.\ Rev.\ D {\bf 53}, 476 (1996)
[arXiv:hep-lat/9507004];\\
C.-J.~D.~Lin {\it et al.}, arXiv:hep-lat/0211043.


\end{thebibliography}
\end{document}